\pdfoutput=1
\documentclass[twocolumn,nofootinbib]{revtex4-1}
\usepackage[utf8]{inputenc}
\usepackage[T1]{fontenc}
\usepackage[english]{babel}
\usepackage{graphicx,amssymb,amsmath,hyperref,rotating}

\newcommand{\nside}{N_\textrm{side}}
\newcommand{\citey}[1]{ \citep{#1}}
\newcommand{\foot}[1]{ \footnote{#1}}

\begin{document}

\title{Lensing simulations by Taylor expansion -- not so inefficient after all}
\author{Sigurd K. Næss}
\email{Sigurd.Naess@astro.ox.ac.uk}
\author{Thibaut Louis}
\affiliation{Sub-department of Astrophysics, University of Oxford, Keble Road, Oxford, OX1 3RH, UK}

\begin{abstract}
	Cosmic Microwave Background lensing simulation by Taylor
	expansion has long been considered impractical due to
	slow convergence, but a recent flat-sky implementation
	shows that a simple trick eliminates this problem, making Taylor
	lensing a fast and simple lensing algorithm for the flat sky. Here we
	generalize the method to the full sky, and study its convergence
	and performance relative to a commonly used numerical code, Lenspix,
	with extensive benchmarks of both.
	Compared to the flat sky case, the method takes a speed hit due to
	the slow speed of spherical harmonic transforms compared to fast Fourier
	transforms, resulting in speeds of
	$\frac13$ to $\frac23$ of Lenspix for similar accuracy.
\end{abstract}

\maketitle

\section{Introduction}
On its way from the surface of last scattering to us, the Cosmic Microwave
Background radiation (CMB) is weakly lensed by the gravitational potential
of the intervening large scale structure\citey{lewis:lensreview:2006}. This introduces a non-linear
distortion that was first detected as a correlation between the CMB
and large scale structure using WMAP data \citet{smith:wmaplens:2007,hirata:lens:2008},
and has since been measured drectly in the CMB at progressively increasing sensitivity
by ACT\citey{das:firstlens,Das:2013zf}, SPT\citey{sptlens:2011,sptlens:2012,sptlens:2012b} and Planck\citey{planck1:XVII:lensing}.

As CMB lensing measurements improve, they have the potential to be some
of the cleanest probes of the total matter distribution in the universe,
especially when coupled with galaxy surveys. This will make them
powerful probes for neutrino masses\citey{2006PhR...429..307L} and the
nature of dark matter and dark energy\citey{kendrick:cosmolens:2006}.

On the scales relevant for CMB observations, this lensing is weak enough to
be modeled as a simple displacement in accordance with the Born
approximation\citey{lewis:lensreview:2006},
\begin{align}
	T'(\vec x) &= T(\vec x + \vec\alpha) \\
	\vec P'(\vec x) &= R \vec P(\vec x + \vec\alpha), \label{eq:lenspol}
\end{align}
where $T'$ ($T$) and $\vec P'$ ($\vec P$) are the lensed (unlensed) temperature and polarization
fields; $\vec \alpha = \nabla \phi$ is the displacement field, with $\phi$ being the lensing potential;
and $R$ is a polarization rotation matrix which is very nearly unity\foot{
	This rotation has two contributions. Firstly, the lensing itself introduces a vanishingly
	small rotation of the polarization angle, which is sometimes called ``gravitational
	Faraday-rotation''. Secondly, parallel transport on the sphere results in a much larger
	but still small rotation which is most significant near the poles.}\citey{Lewis:2005tp}.

Conceptually, constructing a lensed map is as simple as reading off values from the unlensed
map at the appropriate positions, but in practice this is complicated by the fact that
computers work with discretely sampled maps. Hence, the value of the map is only known at
some points on the sphere, which will in general not include the displaced position $\vec x + \vec\alpha$.
The problem of lensing a CMB map hence becomes one of \emph{interpolation}.

Several methods for performing this interpolation have been explored in the literature,
including:
\begin{enumerate}
	\item Nearest neighbor remapping, i.e. truncating the offset position to the nearest
		pixel, and using that as the offset value. This leads to severe pixelization errors,
		but for CMB simulations these can be mitigated by generating the map at much higher
		resolution than needed, remapping at this resolution, and then downgrading in the
		end\citey{Lewis:2005tp,lenshat}.
		However working with higher than necessary resolution maps comes with a large performance
		and memory overhead.
	\item Taylor expansion in terms of the displacement $\vec\alpha$:
		\begin{align}
			T(\vec x + \vec\alpha) &= \sum_{n,j\le n}
			\frac{\alpha_\theta^j \alpha_\phi^{n-j}}{j!(n-j)!}
			\left[\partial_\theta^j \partial_\phi^{n-j} T\right](\vec x).
		\end{align}
		While conceptually simple, this algorithm scales with the expansion order $n$ as
		$\mathcal{O}(n^2)$, which coupled
		with its slow convergence makes it inefficient in practice\citey{lewis:chall:2005,Lewis:2005tp}.
	\item Direct evaluation of spherical harmonics at the offset locations, i.e.
		$T(\vec x + \vec \alpha) = \sum_{lm} T_{lm} Y_{lm}(\vec x + \vec \alpha)$, where $T_{lm}$
		are the spherical harmonic coefficients of the unlensed map. Sadly,
		spherical harmonics transformations (SHTs) are very expensive for unevenly
		spaced pixels\citey{Lewis:2005tp}, though the method can be sped up by
		recasting it as a non-equidistant fast Fourier transform on the torus\citey{torus2008}.
	\item Constrained realizations based on the neighboring pixels and the known
		statistical properties of the signal. This is in theory exact, but in practice
		simplifying approximations of the signal correlations are necessary for
		computational tractability\citey{flints2010}.
	\item The use of interpolation-friendly pixelization schemes, such as ECP.
		This over-pixelizes the sphere compared to HEALPix, but the regular grid
		makes it easy to implement many interpolation schemes from standard
		image processing, such as bi-cubic interpolation. It
		can also be used to speed up some of the previously mentioned methods.
		This is the technique used by the
		popular\citey{smith:wmaplens:2007,sptlens:2012,planck1:XVII:lensing}
		Lenspix algorithm\citey{Lewis:2005tp}.
\end{enumerate}

\begin{figure}[htp]
	\begin{center}
		\includegraphics[width=\columnwidth]{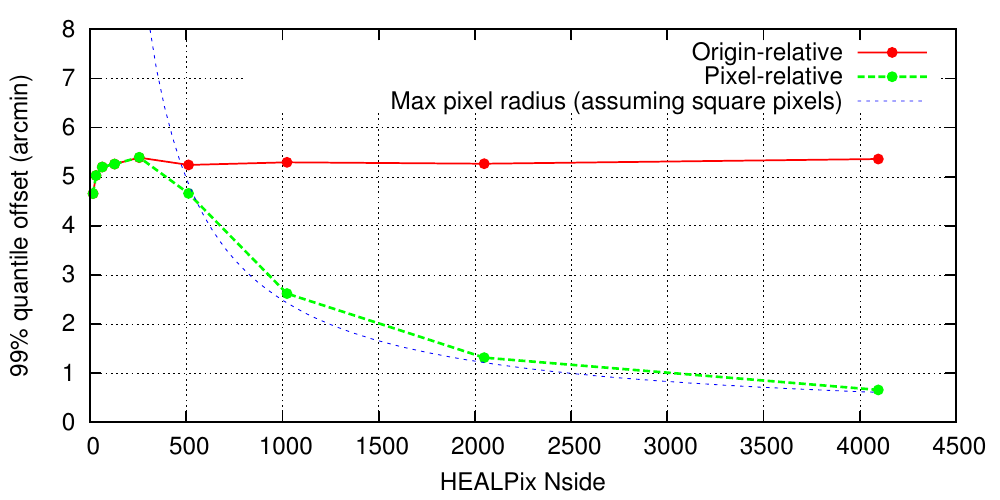}
		\caption{Comparison of the interpolation distance for plain Taylor expansion (red)
			versus expansion relative to the nearest pixel center (green). The 99\% quantile
			of the lensing deflection is 5.3' in our simulations. For $\nside>512$,
			the pixel radius (blue) is smaller than this, making pixel-relative
			expansion increasingly efficient at higher resolution, as a smaller
			expansion parameter means we converge earlier.}
		\label{fig:pixrel}
	\end{center}
\end{figure}
In a recent paper, \citet{louis:lens:2013} showed that a trivial modification of
the Taylor expansion method cures it of its slow convergence, and demonstrated
its excellent accuracy and speed for flat-sky lensing simulations.

In general, the Taylor expansion of a function $f(x)$ around a point $x_0$
becomes less accurate as the distance from $x_0$ grows, and conversely, the expansion can
be truncated earlier if one can expand around a point close to where one wishes to evaluate
the function.

The Taylor expansion used in the literature\citey{Lewis:2005tp,okamoto:lensest:2003,planck1:XVII:lensing}
expands $T(\vec x + \vec\alpha)$ around the point
$\vec\alpha=0$, and the reason for the slow convergence is that $\vec\alpha$ can be relatively
large compared to the scales involved in the map. A better choice is to expand
around the closest pixel center $\vec \alpha_0$, which
is already exactly available, resulting in the following expansion:
\begin{align}
	T'(\vec x) &= T(\vec x + \vec\alpha_0 + \Delta\vec\alpha) \notag \\
		&= \sum_{n,j\le n}
		\frac{\Delta\alpha_\theta^j \Delta\alpha_\phi^{n-j}}{j!(n-j)!}
		\left[\partial_\theta^j \partial_\phi^{n-j} T\right](\vec x + \vec\alpha_0).
\end{align}
At higher resolutions, this results in an expansion in terms of a
much smaller parameter compared to the standard expansion around the undisplaced
location, as shown in figure~\ref{fig:pixrel}.

For the flat sky, this, combined with fast Fourier transforms (FFTs) for computing
the derivatives, results in fast lensed simulations at 0.5 arcminute that are accurate
at all relevant scales by order 2 in the expansion, with no pixel upscaling needed\citey{louis:lens:2013}.

However, this does not necessarily generalize to the full sky. Firstly, the
resolutions relevant for full-sky simulations are much lower. This means that more important
parts of the unlensed power spectra will be truncated. And secondly, SHTs
are much slower than FFTs\foot{Currently practical SHTs scale as $\mathcal(O){N_\textrm{pix}^\frac32}$,
	compared to $\mathcal(O){N_\textrm{pix}\log(N_\textrm{pix})}$ for FFTs, and additionally
the prefactor is much larger for SHTs.}, so computing the derivatives in harmonic space is
much more expensive here than on the flat sky.

The purpose of this paper is to empirically test how important these
issues are in practice by determining the performance of nearest neighbor
Taylor interpolation for full-sky CMB simulations.
\begin{figure*}[htbp]
	\begin{center}
		\includegraphics[width=0.75\textwidth]{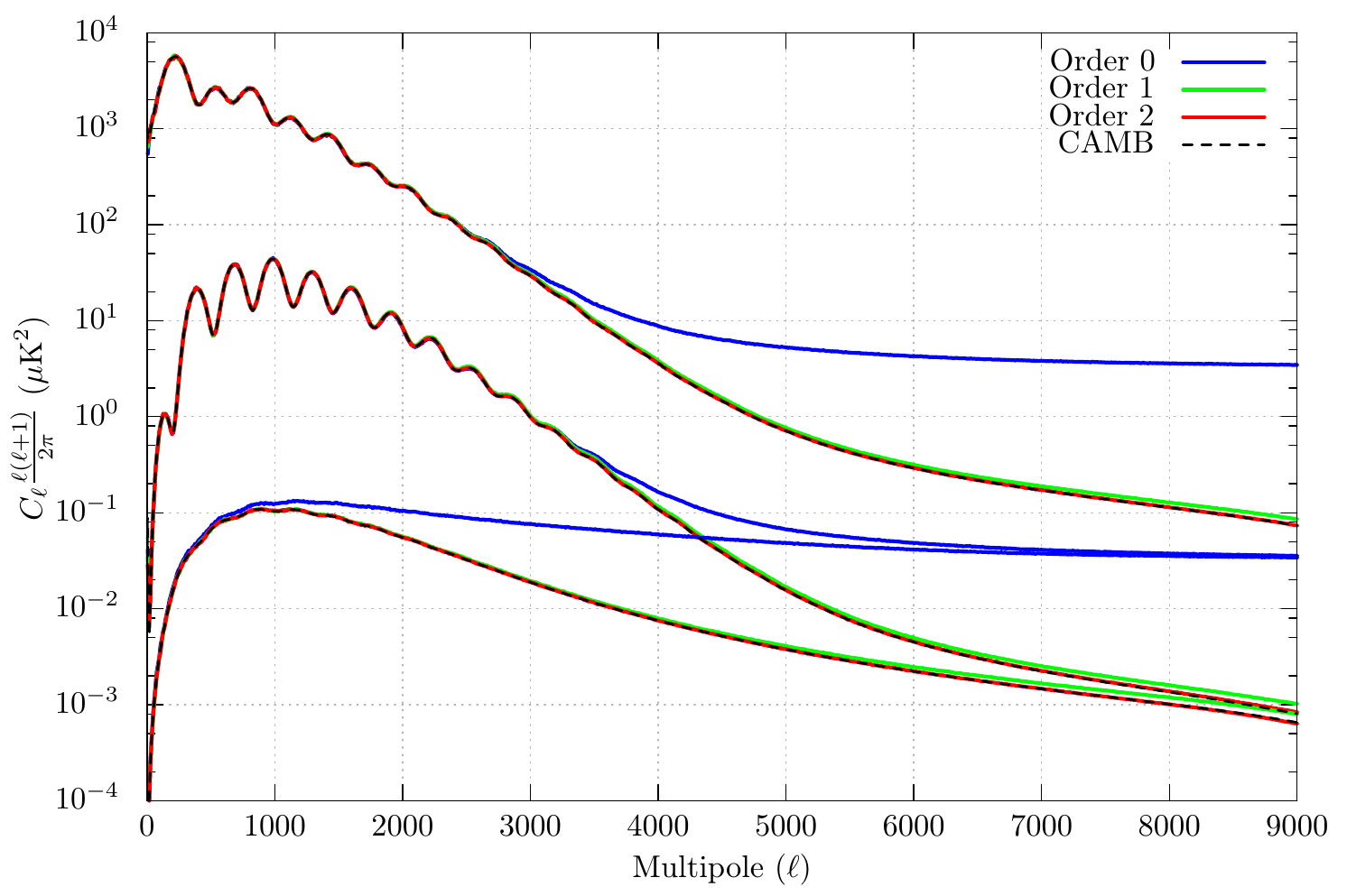}
		\caption{Illustration of the convergence of the pixel-relative Taylor
		expansion at $\nside=4096$ for the TT (top), EE (middle) and BB (bottom)
		spectra. The blue, green and red curves corresponds to stopping the expansion
		at order 0, 1 and 2 respectively. The black curve is the theoretical result
		from CAMB. On this absolute scale, the order 2 curve is indistinguishable
		from theory. See figure~\ref{fig:accuracy} for a more quantitative treatment
		of the convergence.}
		\label{fig:abs_conv}
	\end{center}
\end{figure*}

\section{Implementation}

\subsection{Preparing the input}
The main interpolation step of the algorithm needs a map to lens and a
mapping from lensed to unlensed coordinates that describes the lensing field.
If these are already present from other sources, no further preparation is
needed, and this step can be skipped.

But often one needs to generate lensed maps from scratch, such as for the
accuracy tests in the next section. For this, we simulated HEALPix-pixelated\citey{healpix}
CMB maps (using \verb|healpy.synfast|) and corresponding displacements
$\nabla\phi$ (using \verb|healpy.synalm| and \verb|healpy.alm2map_der1|) with no pixel window.
These were based on theoretical unlensed spectra from CAMB, for a $\Lambda$CDM
cosmological model. For the gradient
calculation we used $\ell_\textrm{max}^{\nabla\phi} = \textrm{min}(8\nside,10\,000)$.
This value is much higher than what is
normally necessary, but is important because the gradient map acts like a high-pass
filter and is very sensitive to small scales. This high $\ell_\textrm{max}^{\nabla\phi}$ comes
at a cost in performance, but it is usually subdominant compared to the other steps,
and we did not attempt to fine-tune $\ell_\textrm{max}^{\nabla\phi}$ for optimal performance.

The offset positions $\vec x' = \vec x + \nabla \phi$ were then computed both
via parallel transport using equations~(A15-A16) in \citey{Lewis:2005tp}
and by the naive method of simply adding the coordinates, which we found to
make no practical difference (see fig.~\ref{fig:norot}).

\subsection{Interpolating}
With an unlensed map $m$ and a position map $\vec x'$ in hand, we can now
calculate the set of nearest pixels
$\{p\}$ using \verb|healpy.ang2pix|, their center locations $\vec x_0'$
via \verb|healpy.pix2ang|, and the subpixel offset $\Delta\vec\alpha = \vec x' - \vec x_0'$.\foot{
	As \texttt{healpy.alm2map\_der1} actually returns $\left(\frac{\partial m}{\partial\theta},
	\frac{\partial m}{\sin\theta\partial\phi}\right)$, the expansion we performed in practice was
	in terms of $\left(\Delta\alpha_\theta,\sin\theta\Delta\alpha_\phi\right)$ to avoid needing
	to rescale all the derivatives.}
We then Taylor-expand each component $s \in \{T,Q,U\}$ of the map $m$ independently to
estimate the value $T'(\vec x)$ and $P'(\vec x)$:
\begin{align}
	s'(\vec x) &= \sum_{n,j\le n} \frac{\Delta\alpha_\theta^j \Delta\alpha_\phi^{n-j}}{j!(n-j)!}
	\left[\partial_\theta^j \partial_\phi^{n-j} s\right](\vec x_0'),
\end{align}
where the higher derivatives were computed by repeated applications of \verb|healpy.map2alm|\foot{
We found that using \texttt{iter=0} led to no appreciable loss in accuracy while giving
a great increase in speed compared to the default \texttt{iter=3}.}
and \verb|healpy.alm2map_der1|.

Using the fact that \verb|healpy.alm2map_der1| returns two derivatives at the same
time, only $\lfloor\frac{n}{2}\rfloor+1$ forward and backward SHTs are needed for each order
$n>0$ in the expansion, for a total of $N+\lfloor\frac{N}{2}\rfloor\left(\lfloor\frac{N+1}{2}\rfloor+1\right)$
two-way SHTs when stopping at order $N$.

Finally, the lensed polarization map can be corrected for the effect of rotation
(i.e. the matrix $R$ in eq.~\ref{eq:lenspol}) due to parallel transport using equations~(A17-A18) in
\citey{Lewis:2005tp},
though we found this to have no appreciable effect on the power spectrum at any
any of the scales we tested.

A small python library implementing this algorithm, called Taylens, is available at \citey{taylens:code}.

\section{Results}
\subsection{Accuracy and convergence}
We performed two sets of tests of the algorithm. In the first, we
systematically investigated accuracy and convergence by
generating a large number of simulations (256 for all but the highest
resolution) for each $\nside \in \{512,\allowbreak1024,\allowbreak2048,\allowbreak4096\}$ and for each expansion
order from 0 to 4. We here used the HEALPix default $\ell_\textrm{max} = 3\nside$
as the accuracy of the calculation of the derivatives in the expansion (but not
when generating the deflected positions). The power spectra of the
resulting maps were computed using \verb|healpy.anafast|, and were
then compared with the result of a high-accuracy CAMB simulation.

\begin{figure*}[ht]
	\begin{center}
		\begin{tabular}{cccc}
			\begin{sideways}\parbox{3.8cm}{\centering\scriptsize Deviation from CAMB (\%)}\end{sideways} &
			\includegraphics[width=0.32\textwidth]{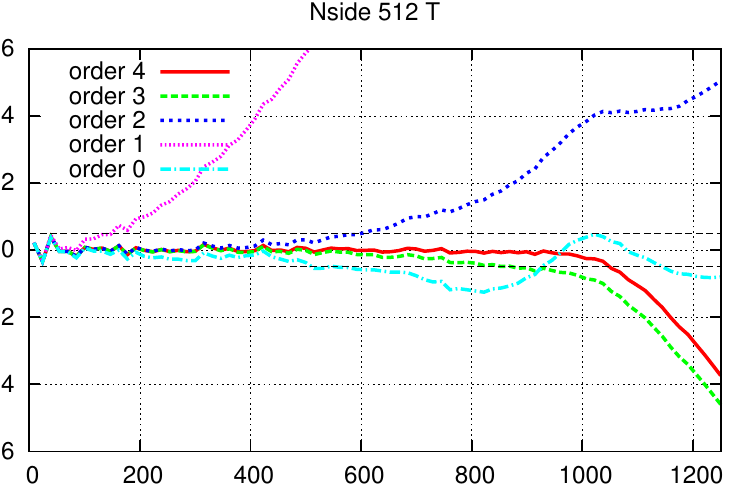} &
			\includegraphics[width=0.32\textwidth]{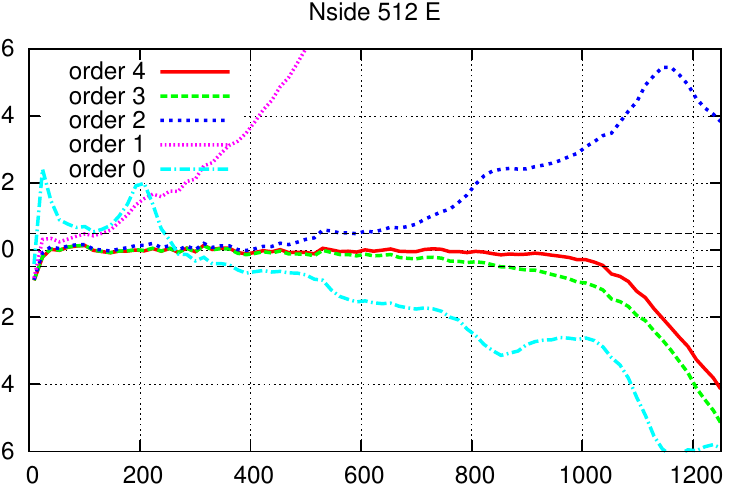} &
			\includegraphics[width=0.32\textwidth]{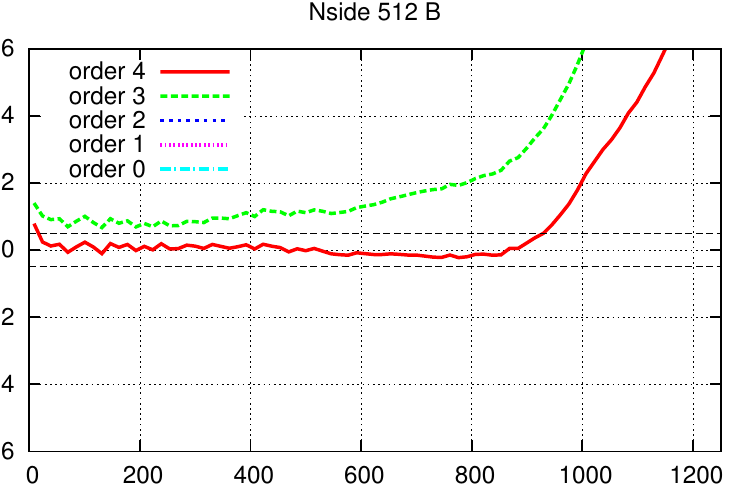} \\
			\begin{sideways}\parbox{3.8cm}{\centering\scriptsize Deviation from CAMB (\%)}\end{sideways} &
			\includegraphics[width=0.32\textwidth]{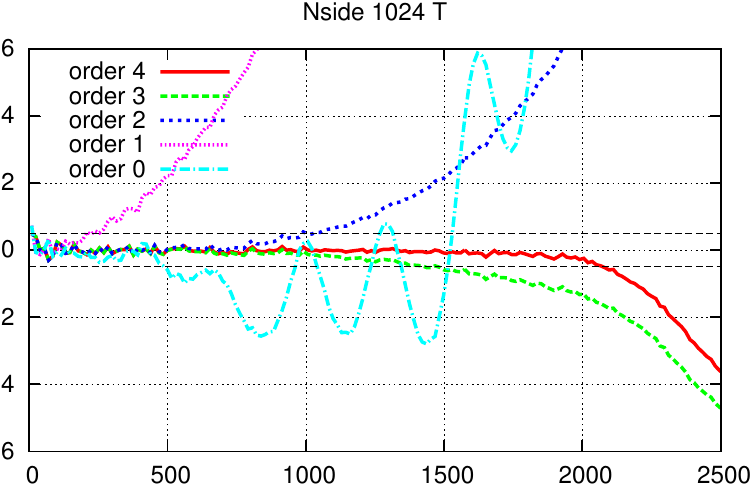} &
			\includegraphics[width=0.32\textwidth]{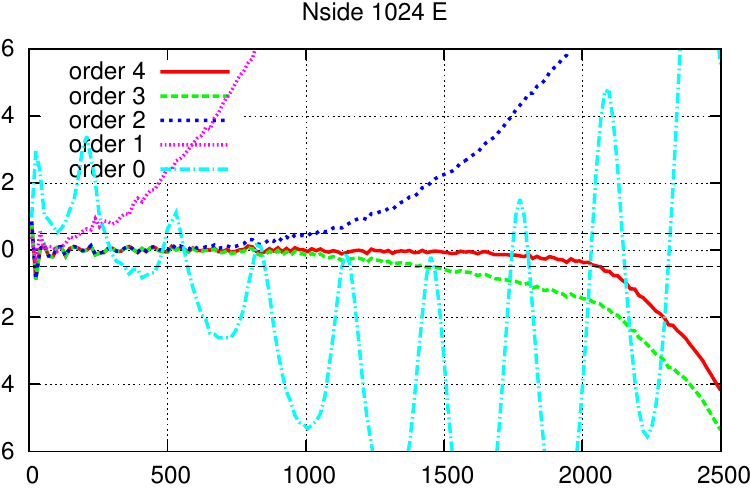} &
			\includegraphics[width=0.32\textwidth]{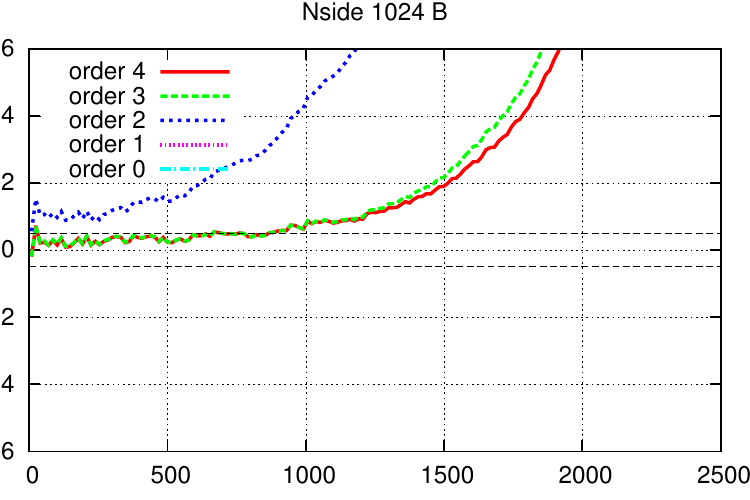} \\
			\begin{sideways}\parbox{3.8cm}{\centering\scriptsize Deviation from CAMB (\%)}\end{sideways} &
			\includegraphics[width=0.32\textwidth]{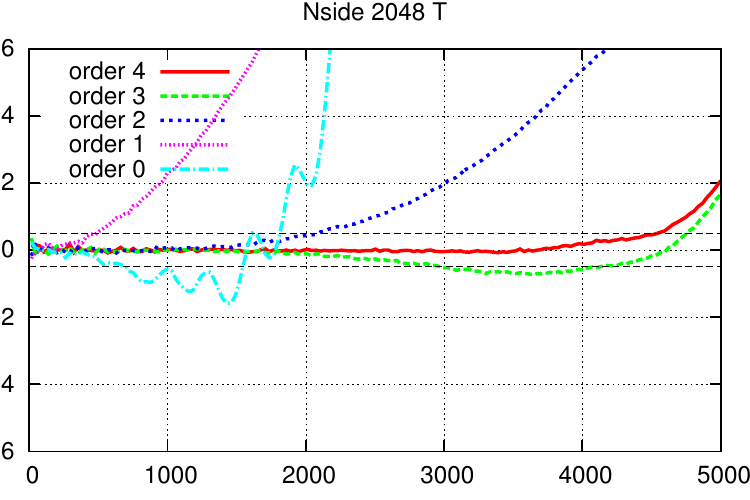} &
			\includegraphics[width=0.32\textwidth]{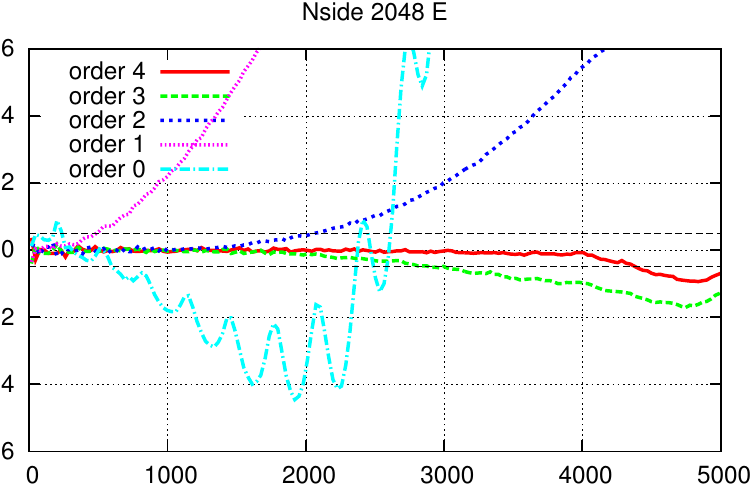} &
			\includegraphics[width=0.32\textwidth]{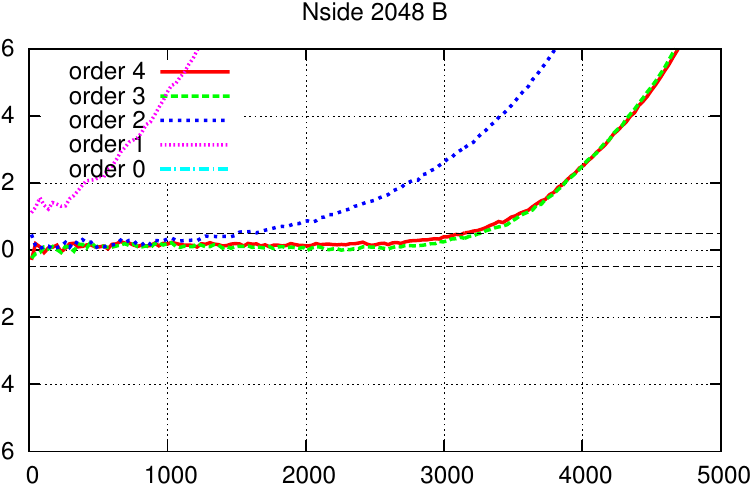} \\
			\begin{sideways}\parbox{3.8cm}{\centering\scriptsize Deviation from CAMB (\%)}\end{sideways} &
			\includegraphics[width=0.32\textwidth]{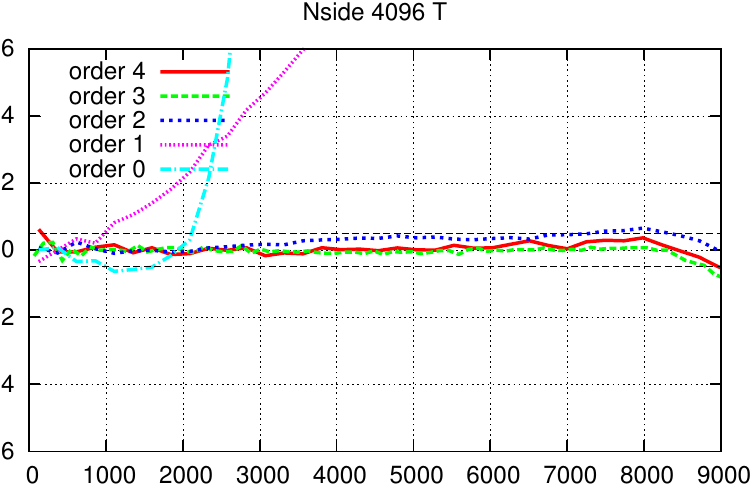} &
			\includegraphics[width=0.32\textwidth]{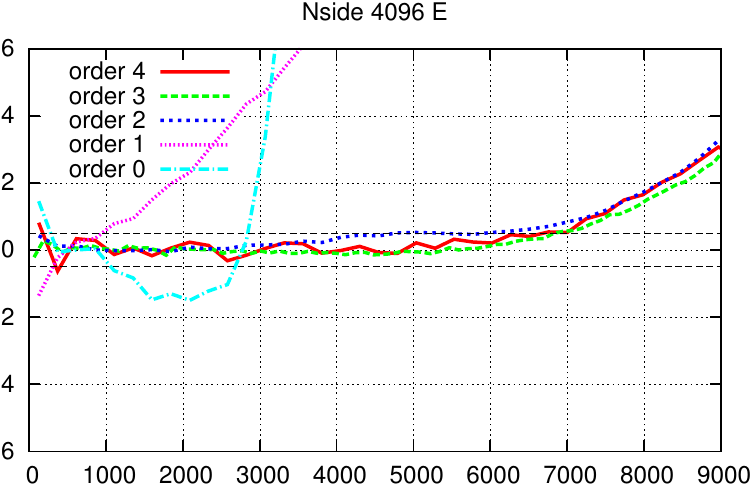} &
			\includegraphics[width=0.32\textwidth]{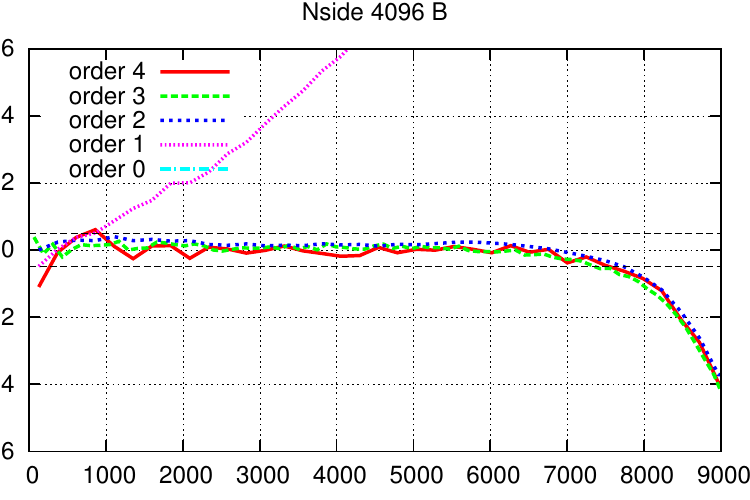} \\
			&Multipole ($\ell$) & Multipole ($\ell$) & Multipole ($\ell$)
		\end{tabular}
		\caption{The relative error (in \%) of Taylor lensing in terms of its output power
			spectra compared to the theoretical CAMB lensed spectra as a function of multipole.
			The three columns are TT, EE and
			BB, while the four rows are simulation $\nside$ of 512, 1024, 2048 and 4096
			respectively. Each panel shows the error for
			truncating the Taylor expansion at orders 0 to 4 inclusive. Typically
			all spectra are accurate to less than 0.5\% up to a multipole of $1.5\nside$ to
			$2\nside$ at order 4. The curves are averages over 256 simulations (with the exception
			of $\nside=4096$, where fewer were used due to high cost).}
		\label{fig:accuracy}
	\end{center}
\end{figure*}

The result can be seen in figures~\ref{fig:abs_conv}  and \ref{fig:accuracy}.
As others have observed,
order 0 (naive nearest neighbor lookup) performs very poorly, and fails to
produce the correct BB power at any scale even at the highest resolutions.
But from order 2 the accuracy quickly improves, and the series is close to
converged by order 4, where it is typically accurate at
the 0.5\% level up to $1.5\nside \lesssim \ell \lesssim 2\nside$. The series
converges faster at higher resolutions, and at $\nside = 4096$ it
is almost fully converged by order 2, where it is accurate up to $\ell \sim 6000$,
which is close to the accuracy limit of CAMB.

\subsection{Benchmarks}
In the second test, we investigated the performance of the algorithm, and compared
it with Lenspix, the most popular lensing simulator at the time of writing.
Lenspix is an example of a local interpolation scheme, and estimates the
value of the CMB at the lensing-displaced positions
by using bi-cubic interpolation after repixlizing the sky in
Equi-Cylindrical coordinates.
Both Lenspix and our HEALPix-based implementation of Taylor lensing (Taylens)
have 3 parameters that affect accuracy and speed: The resolution $\nside$,
the highest multipole $\ell_\textrm{max}$ used in SHTs, and the interpolation
order. In order to test each algorithm at their best, we tested every combination
of $\nside \in \{256,\allowbreak512,\allowbreak1024,\allowbreak2048,\allowbreak4096\}$, $\ell_\textrm{max} \in
\{\nside,\allowbreak\frac32\nside,\allowbreak2\nside,\allowbreak\frac52\nside,\allowbreak3\nside\}$ and $\textrm{order} \in \{0,\allowbreak1,\allowbreak2,\allowbreak3,\allowbreak4\}$
for Taylens and $\textrm{order} \in \{1.0,\allowbreak1.5,\allowbreak2.0,\allowbreak2.5,\allowbreak3.0\}$ for Lenspix, for
a total of 150 combinations for each. Due to computational constraints, only
1 simulation was performed for each of these.

The benchmarks were performed on 8 core 2.53 GHz Xenon E5540 nodes on the SciNet
cluster. All reported times are total CPU times added up across all 8 cores
unless noted otherwise.

\begin{figure}[ht]
	\begin{center}
		\includegraphics[width=\columnwidth]{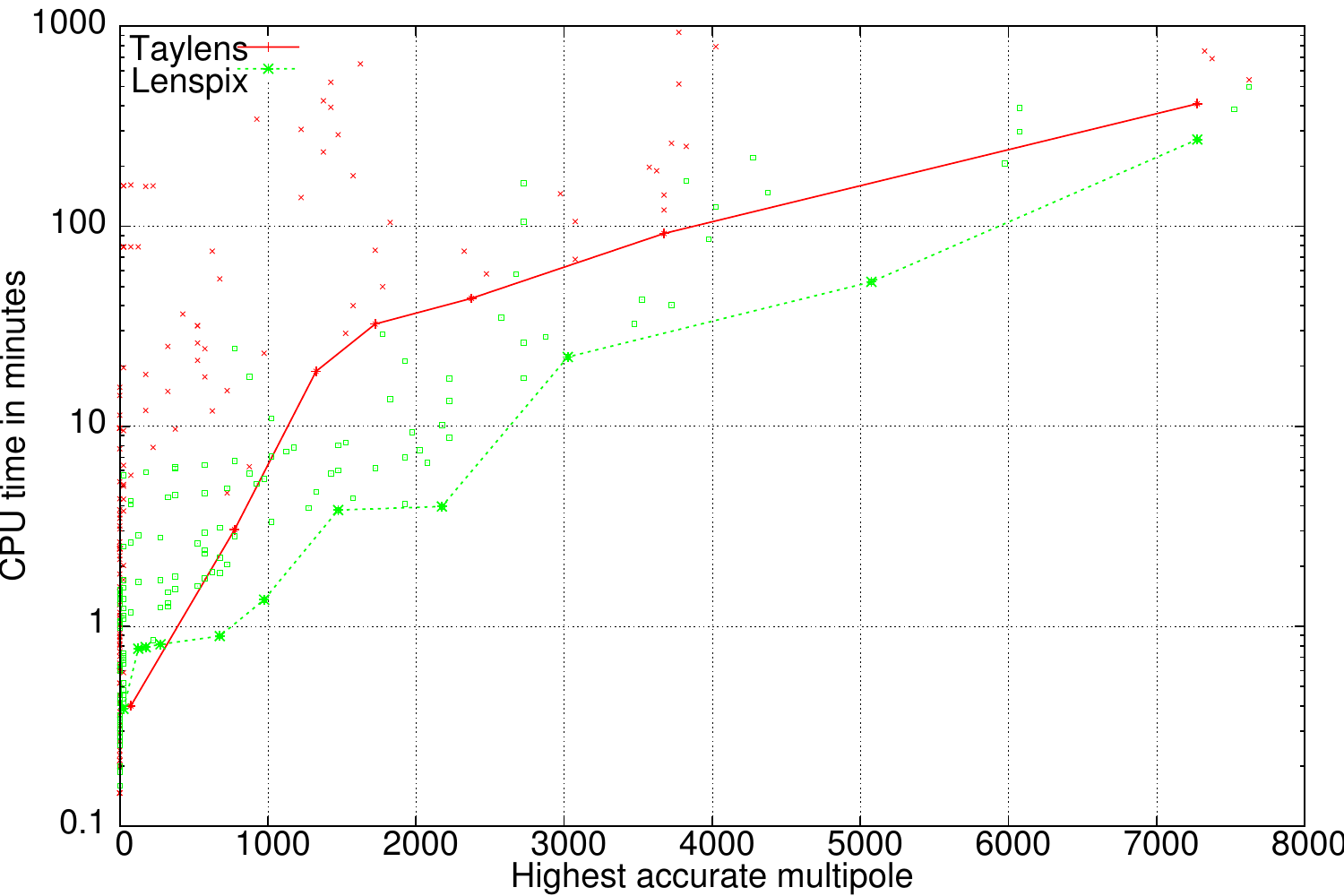}
		\caption{Comparison of the performance of a python implementation of the Taylor
			lensing algorithm described in this paper (Taylens, red, crosses/solid line) with that of
			the popular Lenspix (green, boxes, dashed line).
			Only the interpolation step (which is where they differ) is included in the times.
			To find the optimal parameters for each of these,
			parameters were systematically varied, resulting in the set of small red
			and green points. The lower envelopes of these sets of points is an
			estimate of the optimal performance, and are shown with lines. The horizontal
			axis indicates the maximum multipole for which TT, EE and BB are all within
			0.5\% of the theoretical prediction CAMB, while the vertical axis is the total CPU time in minutes.
			Taylens is slower than Lenspix by a factor of $\sim3$ for most relevant
			scales.}
			\label{fig:performance}
	\end{center}
\end{figure}

The resulting accuracy vs. CPU time is plotted in figure~\ref{fig:performance},
and the most efficient combinations found are summarized in table~\ref{tab:fastest}.
The highest accurate multipole $\ell^*$ is defined here as the center of the
last $\Delta \ell = 50$ bin of each of the spectra that is not significantly (< $4\sigma$\foot{
The threshold of $4\sigma$ was chosen to be high enough that it would be unlikely to be
reached due to a random fluctuation. $\sigma$ here refers to the uncertainty of the
measured mean in each $\Delta \ell=50$ bin.}) above the 0.5\% accuracy limit.

In general, Taylor lensing is slower than Lenspix by a factor of $\sim 3$ for most
relevant accuracies, but towards the very highest accuracies this factor decreases
to 1.5. Compared to the flat-sky case, Taylor lensing with harmonic derivatives
is hampered by the $\mathcal{O}(N_\textrm{pix}^\frac32)$ scaling of the SHT,
compared to the $\mathcal{O}(N_\textrm{pix}\log N_\textrm{pix})$ scaling of the
flat-sky FFT. However, due to the quicker convergence at higher resolution,
the effective scaling is significantly better than the naive estimate of
$\mathcal{O}(\ell^3)$ one would expect based on SHT scaling, and for the parameter
range probed here, it seems to be closer to $\mathcal{O}(\ell^2)$.

As an example, creating a lensed map accurate up to a multipole of 3000
took 68 minutes (8.5 minutes on 8 cores) using Taylens, and 22 minutes
(2.8 minutes on 8 cores) using Lenspix\foot{Taylens indirectly benefits
from HEALPix's OpenMP parallelization for lensing single maps, and also
uses minimal MPI for lensing several independent maps in parallel.}.

\section{Conclusion}
Lensing simulation by nearest neighbor-relative Taylor expansion is a fast, simple
and accurate lensing method on the flat sky, but for the full sky we find
it to be hampered by the poor scaling of spherical harmonic transforms
compared to fast Fourier transforms. Nevertheless, our extensive
benchmarks show that it performs quite well on the full sky, reaching
$\frac13$ to $\frac23$ of the speed of the highly optimized Lenspix code,
while being very simple to implement\foot{Our demonstration implementation
	consists of 253 lines of Python, including comments, I/O and user interface,
	and depends only on healpy and numpy.
}.

The dependence on slow SHTs can be broken by switching to a local pixel-space
estimator for the derivative, though this is somewhat cumbersome on
the curved sky, and may be less accurate than the harmonic derivatives used here.
A better pixel-local approach would probably be the constrained realization
lensing used in FLINTS\citey{flints2010}.

Alternatively, switching to ECP pixelization would make FFTs available,
for a potentially large speed-up (though at the cost of extra projection
operations if one still wishes the output to be in standard HEALPix coordinates).

\begin{acknowledgments}
The authors would like to thank Johannes Noller and Jo Dunkley for useful
discussion. Computations were performed on the gpc supercomputer at
the SciNet HPC Consortium. SciNet is funded by: the Canada Foundation for
Innovation under the auspices of Compute Canada; the Government of Ontario;
Ontario Research Fund - Research Excellence; and the University of Toronto.
SN and TL are supported by ERC grant 259505.
\end{acknowledgments}

\begin{figure}[htp]
	\begin{center}
		\includegraphics[width=\columnwidth]{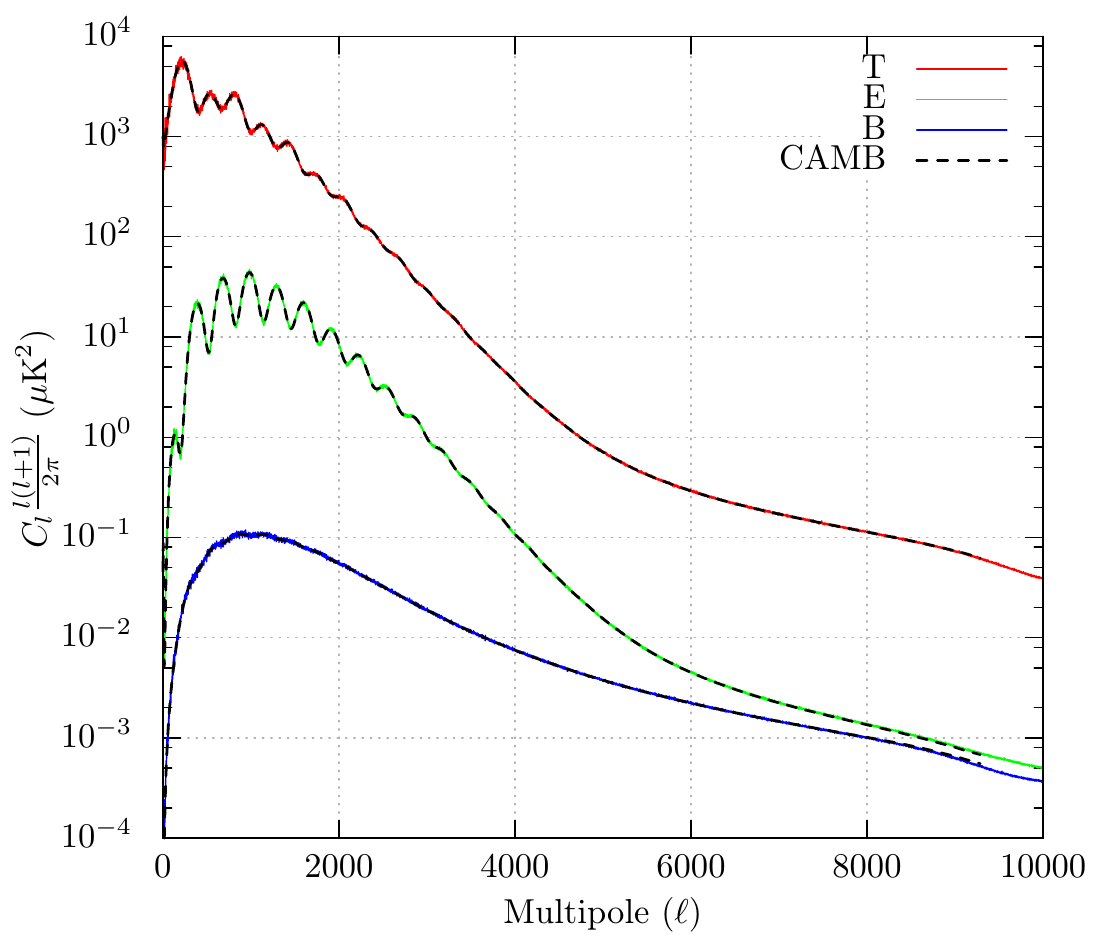}
		\caption{Comparison of the $\nside=4096$ order 3 Taylens spectra (colored curves)
		with the CAMB lensed spectra (dashed black curves).}
		\label{fig:abs}
	\end{center}
\end{figure}

\appendix*
\section{Performance details}
Table~\ref{tab:fastest} shows the most cost-efficient parameters found
during the brute force parameter benchmarks, defined as the parameters
making up the lower-time envelope of the accuracy vs. time distribution.
$\nside \sim \frac23 \ell^*$, $\ell_\textrm{max}
\sim \ell^*+2000$ seems to be a good default choice for both Taylens and Lenspix.

As can be seen from figure~\ref{fig:performance}, the impact of parameter
choice is large -- roughly twice as large as the performance
difference between Taylens and Lenspix for the parameters we explored.

\begin{table}[htp]
	\vspace{1mm}
	\begin{ruledtabular}
		{\bf Taylens}
		\begin{tabular}{rrrrrrr}
		$\ell_{TT}^*$ & $\ell_{EE}^*$ & $\ell_{BB}^*$ & $\nside$ & $\ell_\textrm{max}$ & Order & Time \\
			\hline
			  625& 575&  75& 256 & 768&2&  0.4 \\
			 1075&1075& 775& 512 &1536&3&  3.0 \\
			 2075&1875&1325&1024 &3072&3& 18.8 \\
			 2575&2625&1725&2048 &3072&2& 32.5 \\
			 2725&2675&2375&2048 &4096&2& 43.6 \\
			 4025&3925&3075&2048 &4096&3& 68.4 \\
			 4975&4175&3675&2048 &5120&3& 92.0 \\
			 9275&7275&8175&4096 &6144 $^\textrm{\ref{foot:lmax}}$&2&409.9
		\end{tabular}
		\vspace{1mm}
		{\bf Lenspix}
		\begin{tabular}{rrrrrrr}
		$\ell_{TT}^*$ & $\ell_{EE}^*$ & $\ell_{BB}^*$ & $\nside$ & $\ell_\textrm{max}$ & Order & Time \\
			\hline
			 525& 375&  25&  512& 1024&1.0&  0.4 \\
			1175&1125& 975&  512& 1536&2.0&  1.4 \\
			2175&1825&1475& 1024& 2457&1.5&  3.8 \\
			2175&2225&2175& 1024& 2560&1.5&  4.0 \\
			3025&3025&3725& 2048& 5120&1.5& 22.2 \\
			5525&5275&5075& 2048& 6144&2.5& 52.8 \\
			9275&7275&7625& 4096&10000&1.5&271.7
		\end{tabular}
		\caption{Table of the settings and results for the Taylens (top) and Lenspix (bottom) lower envelope
		lines in figure~\ref{fig:performance}. These indicate the highest cost efficiency found in the
		parameter search. For both Taylens and Lenspix, $\ell_\textrm{max} \gtrsim \ell_{BB}^* + 2000$ in the range
		$1000 \lesssim \ell_{BB}^* \lesssim 4000$ as a rule of thumb. Times are in CPU minutes.}
		\label{tab:fastest}
	\end{ruledtabular}
\end{table}
\footnotetext{Lensing accuracy to $\ell_{TT}^*>9000$ with $\ell_\textrm{max}=6144$
may seem strange,
but recall that the SHTs where $\ell_\textrm{max}$ applies only are used for
computing the derivatives in the expansion, and hence do not affect the 0th order term.
\label{foot:lmax}}

\begin{figure*}[htp]
	\begin{center}
		\includegraphics[width=0.48\textwidth]{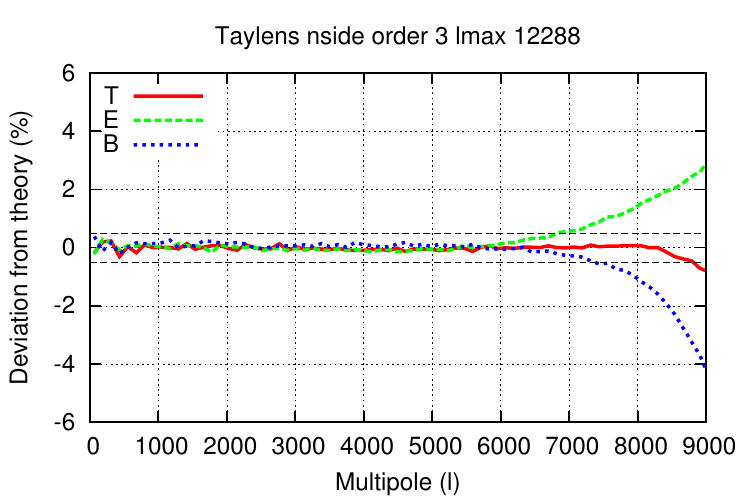}
		\includegraphics[width=0.48\textwidth]{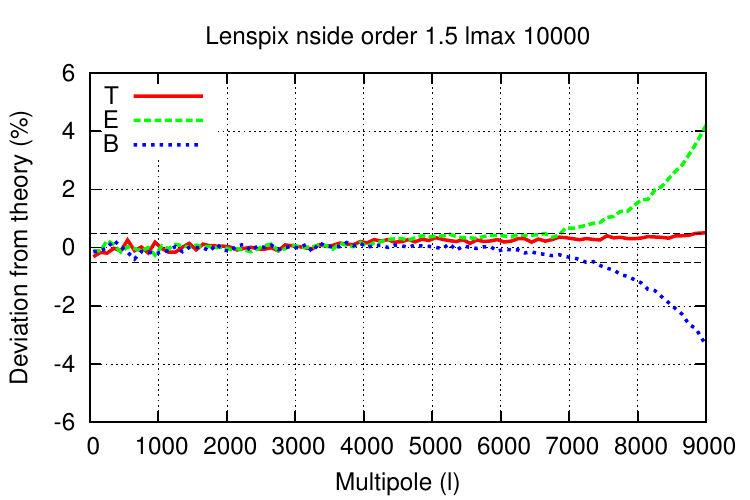}
		\caption{At higher accuracy settings, both Taylens and Lenspix exhibit
		similar behavior at small scales. This starts happening at $\ell\sim 7000$, close
		to the highest multipole CAMB used in the calculation, and we may hence
		be seeing inaccuracy of the CAMB spectra rather than Taylens/Lenspix inaccuracy here.}
		\label{fig:best}
	\end{center}
\end{figure*}

The lensed spectra from Taylens and CAMB are compared on an
absolute scale in figure~\ref{fig:abs}. While the
accuracy information in this graph is the same as (but harder to see than)
figure~\ref{fig:best}, this view shows the large dynamic range over which
the simulations are accurate, which is comparable to that of a 32-bit
floating point number.

At the higher accuracies tested in the benchmarks, both Taylens and Lenspix exhibit
very similar patterns of deviation from CAMB (see figure~\ref{fig:best}), and both
fail to improve further when higher interpolation
orders or $\ell_\textrm{max}$ are used. This points to the error not being in the interpolation
method, but in either the input unlensed power spectrum from CAMB (which is truncated at
$\ell=10\,000$ and not very accurate for $\ell>6000$), or the lensed CAMB power spectrum
that we are comparing with. If so, it is possible that the accuracy of both Taylens and
Lenspix extends beyond $\ell=8\,000-9\,000$ with these settings.

\begin{figure*}[htp]
	\begin{center}
		\includegraphics[height=0.35\textwidth]{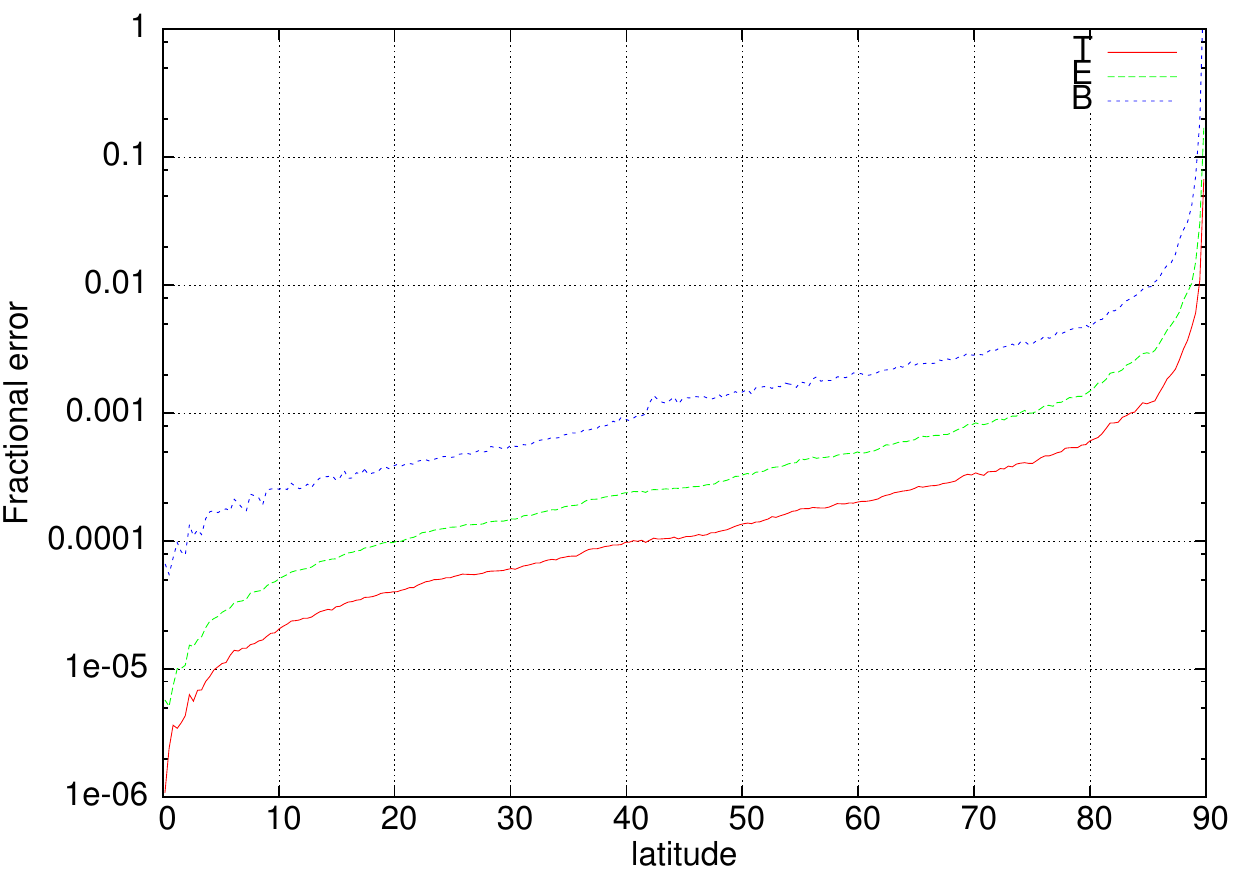}
		\includegraphics[height=0.35\textwidth]{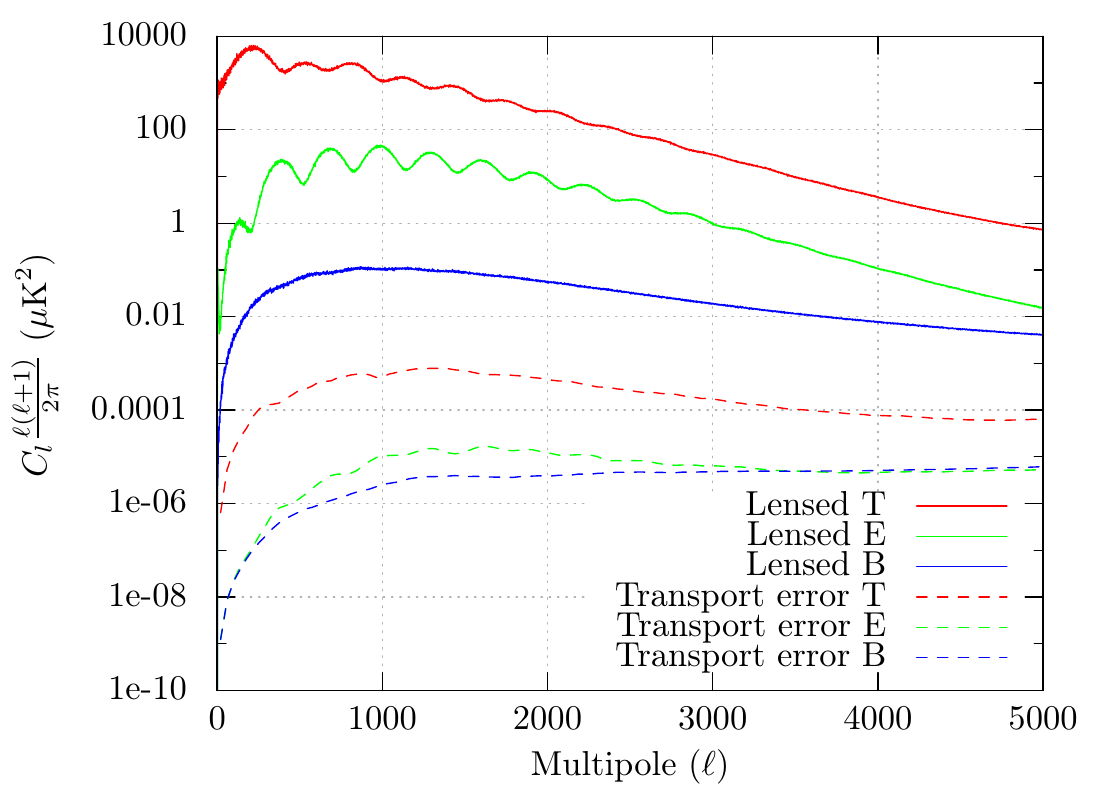}
		\caption{Left panel: The error/signal ratio for the inaccuracy introduced
		when ignoring rotation from parallel transport as a function of latitude.
		Calculated as the standard deviation of the error per latitude bin, divided by
		the signal standard deviation, for each of T, E and B. Excluding the region
		above $85^\circ$, the error is always less than 1\%. Right panel: The
		angular error power spectrum compared with that of the signal, computed for
		$N_\textrm{side}=2048$. The error is much smaller than the signal at all relevant
	scales.}
		\label{fig:norot}
	\end{center}
\end{figure*}

\bibliography{refs}

\end{document}